\def\Version{{5}}






\message{<< Assuming 8.5" x 11" paper >>}    

\magnification=\magstep1	          
\raggedbottom

\parskip=9pt

%

\def\singlespace{\baselineskip=12pt}      
\def\sesquispace{\baselineskip=16pt}      
\def\sesquispace{\baselineskip=13pt}      



\def\ideq{\equiv}		
\def\half{{1 \over 2}}
\def\implies{\Rightarrow}
\def\past   {\mathop {\rm past     }\nolimits}

\font\titlefont=cmb10 scaled\magstep2 
\def\eprint#1{{\tt #1}}
\def\author#1 {\medskip\centerline{\it #1}\bigskip}
\def\address#1{\centerline{\it #1}\smallskip}
\def\furtheraddress#1{\centerline{\it and}\smallskip\centerline{\it #1}\smallskip}
\def\AbstractBegins
{
 \singlespace                                        
 \bigskip\leftskip=1.5truecm\rightskip=1.5truecm     
 \centerline{\bf Abstract}
 \smallskip
 \noindent	
 } 
\def\AbstractEnds
{
 \bigskip\leftskip=0truecm\rightskip=0truecm       
 }

\def\ReferencesBegin
{
 \singlespace					   
 \vskip 0.5truein
 \centerline           {\bf References}
 \par\nobreak
 \medskip
 \noindent
 \parindent=2pt
 \parskip=6pt			
 }

\def\subsection #1 {\medskip\noindent{\subheadfont #1}\par\nobreak\noindent}
\def\reference{\hangindent=1pc\hangafter=1} 
\def\ref{\reference}
\def\sepref{\parskip=4pt \par \hangindent=1pc\hangafter=0}
\def\journaldata#1#2#3#4{{\it #1\/}\phantom{--}{\bf #2$\,$:} $\!$#3 (#4)}

\def\linebreak{\hfil\break}
\def\lbr{\linebreak}
\font\subheadfont=cmssi10 scaled\magstep1 
\def\email#1{\smallskip\centerline{\it address for email: #1}}
\def\PrintVersionNumber{
 \vskip -1 true in \medskip 
 \rightline{version \Version} 
 \vskip 0.3 true in \bigskip \bigskip}


\def\sepref{\parskip=4pt \par \hangindent=1pc\hangafter=0}



\def\webtilde{\lower2pt\hbox{${\widetilde{\phantom{m}}}$}}
\def\webhome{{\tt {http://www.physics.syr.edu/}{\webtilde}{sorkin/}}}




\def\webhomePapers{\webhome{\tt{some.papers/}}}


\fontdimen16\textfont2=2.5pt
\fontdimen17\textfont2=2.5pt
\fontdimen14\textfont2=4.5pt
\fontdimen13\textfont2=4.5pt 

\def\braket#1{\langle#1\rangle}



\phantom{}


\PrintVersionNumber


\sesquispace
\centerline {\titlefont Is the cosmological ``constant'' a nonlocal quantum} 
\centerline{{\titlefont residue of discreteness of the causal set 
              type?}\footnote{$^{^{\displaystyle\star}}$}%
{To appear in the proceedings of the PASCOS-07 Conference, 
  held July, 2007, London, England 
  (American Institute of Physics, to appear online).  
 \eprint{gr-qc/yymmnnn}}}

\bigskip


\singlespace			        

\author{Rafael D. Sorkin}
\address
 {Perimeter Institute, 31 Caroline Street North, Waterloo ON, N2L 2Y5 Canada}
\furtheraddress
 {Department of Physics, Syracuse University, Syracuse, NY 13244-1130, U.S.A.}
\email{sorkin@physics.syr.edu}

\AbstractBegins                              
   The evidence for an accelerating Hubble expansion appears to have
   confirmed the heuristic prediction, from causal set theory, of a
   fluctuating and ``ever-present'' cosmological term in the Einstein
   equations.  A more concrete phenomenological model incorporating this
   prediction has been devised and tested, but it remains incomplete.  I
   will review these developments and also mention a possible
   consequence for the dimensionality of spacetime.
\AbstractEnds


\sesquispace
\vskip -30pt

\subsection{}   

  The inference from causet theory of a fluctuating cosmological
  constant $\Lambda$ is possibly the earliest theoretical prediction of a
  non-zero $\Lambda$; and yet relatively little work has been devoted
  to developing it.  In its original form [1], 
  the prediction yielded
  only an order of magnitude estimate for $\Lambda$, namely that its
  current value should be about $\pm 10^{-120}$ in Planck
  units.  When evidence started accumulating for a $\Lambda$ of just
  this size, it seemed time to try to embed the original
  prediction in a more complete model.  The resulting model, 
  as elaborated
  by Scott Dodelson, Patrick Greene, Maqbool Ahmed and me, had
  some ad hoc elements, but it realized concretely the original
  implication of an ``ever-present $\Lambda$'', that is one whose
  (fluctuating) magnitude is always comparable to the
  Hubble scale. [2] [3]

  My talk today will review these developments in the hope that
  people will be encouraged to carry the underlying idea further.
  I will explain where the original heuristic prediction came
  from, and I will describe a concrete model that arose from
  it. Before doing so, however I want to mention the two greatest
  weaknesses that the model had.

  The first weakness was a certain arbitrariness in how one
  interprets the idea of a varying $\Lambda$. The second was the
  imposition of spatial homogeneity and isotropy on the model, ie
  the assumption that the metric takes the FRW form.  The first
  weakness has been largely overcome in subsequent work by Maqbool
  Ahmed but the second still needs to be tackled.

\subsection{Lambda forgotten} 
The cosmological constant originated as a non-solution to a non-problem,
{\it viz} the fact that the Einstein equation 
$$
  {1\over\kappa} G^{ab} + \Lambda g^{ab} = T^{ab}  \eqno(1)
$$
does not admit a static cosmos as a solution if the $\Lambda$-term is
omitted.  This, of course, was not a real problem because the cosmos is
not static.  Nor does the inclusion of $\Lambda$ solve this non-problem,
because the resulting static cosmos is unstable to collapse or unbounded
expansion, as is well known. 
In view of this inauspicious beginning, the introduction of $\Lambda$ must have
seemed unmotivated, and the prejudice grew up among cosmologists that
$\Lambda$ was 0.  With few exceptions, they either set $\Lambda=0$, or
they never even bothered to mention it at all, despite many indications
from quantum field theory and quantum gravity that things could not
possibly be that simple.

\subsection{Lambda remembered} 
Within the last decade or so, all that has changed, and the origin of
the so called ``dark energy'' is recognized as a central question of
astronomy.  Although many lines of thought and observation seem to have
contributed to the change (including, for example, the problem that
without $\Lambda$ (or with $\Lambda<0$), the cosmos seemed to be younger
than some of its contents),
the two most persuasive arguments concerned the CMB relic
radiation\footnote{$^\star$} 
{CMB = cosmic microwave background radiation.  It was already detected
 as early as 1940 [4], but its significance was not
 appreciated, and it was forgotten.}
and the observations of distant supernovas.  Let us
briefly recall both of these arguments.

\subsection {Spatial flatness and missing mass}
As I understand it, this argument proceeds from the angular size of the
so called ``acoustic peaks'' in the CMB to the spatial flatness of the universe
(in our vicinity) to the need for a new term in the Einstein
equation.  The density and temperature of the medium that emitted the
CMB photons resulted from the initial gravitational collapse and
subsequent fluid oscillation of that medium.  This process produced
bright regions of approximately known diameter and distance from us,
which correspond to the peaks one sees in plots of CMB
brightness vs. angular size.  But since the {\it apparent} angular size
of a distant object is greater in spherical space than in flat space
(and correspondingly less in hyperbolic space), we can infer the radius of curvature of
our cosmos, assuming it to be spatially homogeneous and isotropic (Friedmann
cosmos).  The best fit is to zero curvature, i.e. to spatial flatness.
But for such a cosmos, the $a=b=0$ component of eq. (1) reads, if we
take $\Lambda=0$,
$$
    3 \left({\dot{a}\over a}\right)^2 = \rho  \eqno(2)
$$
where $\rho=T^{00}$ is the total energy density in matter (including the
so-called dark matter inferred from galactic rotation curves and
gravitational lensing).  Also, $a$ is the scale-factor or ``radius of
the universe'', $\dot{a}=da/d\tau$ with $\tau$ = proper time, and I've
set $\kappa\ideq8\pi G = 1$.
By comparing this equality with various, more or less direct
measurements of $\rho$,  
one finds that the latter
would have to be tripled in order to satisfy (2).
With $\Lambda$ restored on the other hand, one obtains instead of
(2) the equation 
$$
      3 \left({\dot{a}\over a}\right)^2 - \Lambda = \rho  \eqno(3)
$$
from which $\Lambda$ can be determined once $\rho$ and $H=\dot{a}/a$
are known.
The conclusion seems to be that either $\Lambda$ is nonzero or
something rather like it is out there, carrying
twice the effective mass-density of non-gravitational matter.

\subsection {Dim distant supernovas}
Observations of supernovas (of type IA) have yielded the most direct
evidence for a positive $\Lambda$, because they let us deduce the value
of the ``acceleration'' $\ddot{a}$ from a plot of luminosity
vs. redshift, and the analysis  depends only on the well
understood behavior of electromagnetic fields in curved spacetime. 
This yields for the ``luminosity distance'' or normalized ``dimness'' 
$d_L$ of
a known source that is not too far away, the equation 
$$
    H d_L = z + \half \left( 1 + {a \ddot{a} \over \dot{a}\dot{a}} \right)
    z^2 + O(z^3)
    \eqno(4)
$$ 
where $H=\dot{a}/a$ is the Hubble constant and $z$ the redshift of the
light from the supernova.  Clearly, one can deduce both $\dot{a}$ and 
$\ddot{a}$ if one knows how $d_L$ varies with $z$.  Moreover one sees
that, other things being equal, a larger $\ddot{a}$ will produce an
image that is {\it dimmer} at equal redshift.  The graph of $d_L$
as a function of $z$ thus shifts upward, and this is what has been seen.

Actually, the sign of this effect can be deduced from the equivalence
principle, using only what we know of the Doppler shift in flat
spacetime.
According to the equivalence principle, spacetime {\it is} flat in our
neighborhood, whence what is usually described as the
expansion of space  we can reinterpret locally as the flow of galaxies away from us.
In such a coordinate system, two supernovas of equal brightness are at
equal spatial distance from us.  Hence their light was emitted at the
same time $t_0$ in the past.  Now let these two stars be in two
different spacetimes with the same expansion rate $H$ but different
accelerations $\ddot{a}$.  In the cosmos with {\it greater}
acceleration, the expansion rate at $t_0$ was smaller, whence the
recession rate of the supernova was smaller, whence its redshift $z$ was
also less.  Equivalently, the supernova appears dimmer at the same
redshift.  

To complete the argument, observe that a positive $\Lambda$ serves to
increase $\ddot{a}$.  This follows from the $a=b=1$ component of
(1), and can be remembered from the fact that, for pure $\Lambda>0$,
one obtains a de Sitter cosmos, in which $a$ grows exponentially with
proper time $\tau$.
  By
contrast, in a cosmos with $\Lambda=0$, the expansion necessarily slows
down, responding to the gravitational attraction of the galaxies for
each other.  

\subsection {The $\Lambda$ puzzle}
The evidence we have just reviewed points to a cosmological constant of
magnitude, $\Lambda\approx 10^{-120}\kappa^{-2}$, and this raises two
puzzles:\footnote{$^\dagger$}
{I prefer the word puzzle or riddle to the word problem, which suggests
 an inconsistency, rather than merely an unexplained feature of our
 theoretical picture.}
Why is $\Lambda$ so small without vanishing entirely, and Why is it so
near to the critical density $\rho_{critical}=3H^2$ that appears in
eqs. (2) and (3)?  Is the latter just a momentary occurrence
in the history of the universe (which we are lucky enough to witness),
or has it a deeper meaning?
 
Clearly both puzzles would be resolved if we had reason to believe that 
$\Lambda\approx H^2$ always.  In that case, the smallness of $\Lambda$
today
would merely reflect the large age of the cosmos.  But such a $\Lambda$
would conflict with our present understanding of nucleosynthesis in the
early universe and of ``structure formation''  more recently. 
(In the first case, the problem is that the cosmic expansion rate
influences the speed with which the temperature falls through the
``window'' for synthesizing the light nuclei, and thereby affects their
abundances.  According to (3) a positive $\Lambda$ at that time
would have increased the expansion rate, which however is already
somewhat too big to match the observed abundances.  In the second case,
the problem is that a more rapid expansion during the time of
structure formation would tend to oppose the enhancement of
density perturbations due to gravitational attraction, making it
difficult for  galaxies to form.)
But neither of these reasons excludes a {\it fluctuating} $\Lambda$ with
typical magnitude
$|\Lambda|\sim H^2$ but mean value $\braket{\Lambda}=0$.  
The point now is that {\it such fluctuations can arise as a residual,
nonlocal quantum effect of discreteness}, and specifically of the type
of discreteness embodied in the causal set.

\subsection{Features of causet theory needed in the following}
In order to explain this claim, I will need to review some basic aspects
of causet theory.  [5]
According to the causal set hypothesis, the smooth
manifold of general relativity dissolves, near the Planck scale, into a
discrete structure whose {\it elements} can be thought of as the ``atoms
of spacetime''.  These atoms can in turn be thought of as representing
``births'', and as such, they carry a relation of ancestry that
mathematically defines a {\it partial order}, $x\prec y$.  
Moreover, in our best dynamical models [6],
the births happen sequentially
in such a way that the number $n$ of elements plays the role of an
auxiliary time-parameter.  (In symbols, $n\sim t$.)\footnote{$^\flat$}
{It is an important constraint on the theory that this auxiliary
 time-label $n$ should be ``pure gauge'' to the  extent that it
 fails to be determined by the physical order-relation $\prec$.  That
 is, it must not influence the dynamics, this being the discrete analog
 of general covariance.}

Two basic assumptions complete the kinematic part of the story by
letting us connect up a causet with a continuum spacetime.  
One posits first,
that the underlying microscopic order $\prec$ corresponds to the
macroscopic relation of before and after, and second, that the number
of elements $N$ comprising a region of spacetime equals the {\it volume}
of that region in fundamental (i.e. Planckian) units.
(In slogan form: geometry = order +  number.)
The equality between number $N$ and volume $V$ is not precise however,
but subject to Poisson fluctuations, whence instead of $N=V$, 
we can write only
$$
       N \sim V \pm \sqrt{V}  \ .    \eqno(5)
$$
(These fluctuations express a ``kinematical randomness'' that seems to
be forced on the theory by the noncompact character of the Lorentz
group.)

To complete the causet story, 
one must provide a
``dynamical law'' governing the birth process by which the causet
``grows'' (the discrete counterpart of equation (1)).
This we still lack in its quantum form, but for heuristic purposes we
can be guided by the classical sequential growth (CSG) models
referred to above; and this
is what I have done in identifying $n$ as a kind of time-parameter.

\subsection{Ever-present $\Lambda$}
We can now appreciate why one might expect a theory of quantum gravity
based on causal sets to lead to a fluctuating cosmological constant.
Let us assume that at sufficiently large scales the effective theory of
spacetime structure is governed by a gravitational path-integral, which
at a deeper level will of course be a sum over causets.  That $n$ plays
the role of time in this sum suggests that it must be held fixed, which
according to (5) corresponds to holding $V$ fixed in the integral
over 4-geometries.  If we were to fix $V$ exactly, we'd be doing
``unimodular gravity'', in which setting it is easy to see that $V$ and
$\Lambda$ are conjugate to each other in the same sense as energy and
time are conjugate in nonrelativistic quantum mechanics.  
[This conjugacy shows up
most obviously in the $\Lambda$-term in the gravitational
action-integral, which is simply
$$
          - \Lambda \int \sqrt{-g} \; d^4x = - \Lambda V \ . \eqno(6)
$$
It can also be recognized in canonical formulations of unimodular
gravity [7], and in the fact that (owing to
(6)) the ``wave function'' $\Psi(^3g;\Lambda)$ produced by the
unrestricted path-integral with parameter $\Lambda$ is just the Fourier
transform of the wave function $\Psi(^3g;V)$ produced at fixed $V$.]
In analogy to the $\Delta{E}\Delta{t}$ uncertainty relation, we thus
expect in quantum gravity to obtain
$$
     \Delta \Lambda  \; \Delta V \sim \hbar \eqno(7)
$$
Remember now, that even with $N$ held exactly constant, $V$ still
fluctuates, following (5), between $N+\sqrt{N}$ and $N-\sqrt{N}$;
that is, we have 
$N\sim V\pm\sqrt{N} \implies V \sim N \pm \sqrt{V}$, or
$\Delta{V}\sim\sqrt{V}$.  In combination with (7), this yields 
for the fluctuations in $\Lambda$ 
the central result 
$$
         \Delta \Lambda  \sim V^{-1/2}  \eqno(8)
$$
Finally, let us {\it assume} that, for reasons still to be discovered,
the value about which $\Lambda$ fluctuates is strictly zero:
$\braket{\Lambda}=0$.
(This is the part of the $\Lambda$ puzzle we are {\it not} trying to
solve.\footnote{$^\star$}
{But see the ansatz (10) below, which yields ${\braket{\Lambda}}=0$
 automatically.}
) 
A rough and ready estimate identifying spacetime volume with the Hubble
scale $H^{-1}$ then yields
$$
   V \sim (H^{-1})^4 \sim H^{-4} \  \implies\ \Lambda \sim V^{-1/2} \sim
   H^2 \sim \rho_{critical}
$$
(where I've used that $\Lambda=\Lambda - \braket{\Lambda}$ since
$\braket{\Lambda}=0$). 
In other words, $\Lambda$ would be ``ever-present'' (at least in 3+1
dimensions).

\subsection{A concrete model incorporating equation (8)}
In trying to develop (8) into a more comprehensive model, we not
only have to decide exactly which spacetime volume `$V$' refers to, we
also need to interpret the idea of a varying $\Lambda$ itself.
Ultimately the phenomenological significance of $V$ and $\Lambda$ would
have to be deduced from a fully developed theory of quantum causets, but
until such a theory is available, the best we can hope for is a 
reasonbly plausible
scheme which realizes (8) in some recognizable form.

As far as $V$ is concerned, it pretty clearly wants to be the volume to the past
of some hypersurface, but which one?  If the local notion of ``effective
$\Lambda$ at $x$'' makes sense, and if we can identify it with the
$\Lambda$ that occurs in (8), then it seems natural to interpret $V$
as the volume of the past of $x$, or equivalently (up to Poisson
fluctuations) as the number of causet elements which are ancestors of
$x$:
$$
              V = {\rm volume}(\past (x))  \ .
$$
One could imagine other interpretations,\footnote{$^\dagger$}
{For example, interpretations in which $\Lambda$ is not a spacetime
 field at all, but must be understood more nonlocally.}
but this seems as simple and direct as any.

As far as $\Lambda$  is concerned, the problems begin with eq. (1)
itself, whose divergence implies (at least naively\footnote{$^\flat$}
{Naively because it neglects the circumstance that a fluctuating $\Lambda$
 would be something like a stochastic Brownian function that need not
 even have a derivative.
})
that $\Lambda$ = constant. 
The model of [2] and [3] addresses this difficulty in two
stages.  First it limits itself to spacetimes of the FRW form, i.e. it
assumes that the cosmos is spatially homogeneous and isotropic
(which of course requires for consistency that $\Lambda$ {\it also} be
spatially homogeneous).
Having assumed this, we might as well assume in addition that 
space is
flat ($k=0$), since that simplifies the equations and matches current
data.  The Einstein equations (1) then reduce (with $\kappa=1$) to a
pair of ordinary differential equations known as the Friedmann
equations: 
$$ 
   3 (\dot{a}/a)^2 = \rho + \rho_\Lambda		\eqno(9a)
$$
$$
   2 \ddot{a}/a +  (\dot{a}/a)^2 = - (p + p_\Lambda) \ , \eqno(9b) 
$$
where $\rho_\Lambda\ideq\Lambda$ and $p_\Lambda\ideq-\Lambda$
(corresponding to the familiar ``equation of state'' of $\Lambda$, $p=-\rho$.)

Now in the usual case where $\Lambda$ is time-independent, equation
(9b) is a consequence of (9a); and conversely the two
equations are incompatible when $\dot{\Lambda}\not=0$, this being
precisely the difficulty with which we began.  To bypass this
incompatibility we are forced to modify the Friedmann
equations.\footnote{$^\star$} 
{Notice in this connection that (9b) will not even be well defined
 if (9a) holds and $\Lambda$ is a function of Brownian motion type.}
The most
straightforward way of doing so is to retain
only one of them, or possibly some other linear combination
of (9a) and (9b).
In reference [2] we followed this approach by adopting
(9a) as our ``dynamical guide'' and discarding (9b).
This choice is appealing because 
the resulting dynamics is easy to simulate, and because it
admits an alternative description in which {\it neither} (9a) nor
(9b) is compromised, but instead the ``equation of state of
$\Lambda$'' is modified in a simple, local manner.
Fortunately, changing one's ``guide'' by adopting a different linear
combination of (9a) and (9b) appears to alter nothing
qualitatively
[3], so let me 
ignore such possibilities for now.
Then our dynamical scheme is just (9) with
$$
\eqalign{
          \rho_\Lambda &= \Lambda  \cr
             p_\Lambda &= - \Lambda - \dot\Lambda / 3H   \cr
}
$$

Finally, to complete our model
and obtain a closed system of equations,
 we need to specify $\Lambda$ as a
(stochastic) function of $V$, 
and we need to choose it so that 
$\Delta\Lambda\sim V^{-\half}$.
But this is actually  easy to accomplish,
if we begin by observing that 
(with $\kappa=\hbar=1$)
$\Lambda=S/V\approx S/N$ can be
interpreted as the action per causet element that is present even when
the spacetime curvature vanishes.
(As one might say, it is the action that an element contributes just by
virtue of its existence.\footnote{$^\dagger$}
{More properly one should probably think of each element as contributing
 a multiplicative {\it phase} $\exp(iS)$.  However, our analysis here is
 only being used to suggest a simple ansatz for $\Lambda$, and this
 ansatz can stand on its own in the present, phenomenological context.})
Now imagine that each element contributes (say) $\pm\hbar$ to $S$, with
a random sign.  Then $S$ is just the sum of $N$ independent random
variables, and we have
$$
           S/\hbar   \sim  \pm\sqrt{N}  \sim  \pm\sqrt{V/\ell^4}  \ ,
$$
where $\ell\sim\sqrt{\hbar\kappa}$ is the fundamental time/length
of the underlying theory, 
which thereby enters our model as a free phenomenological  parameter.
This in turn implies, as desired, that
$$
     \Lambda \ = \  S/V   \ \sim \  \pm \; {\hbar / \ell^2 \over \sqrt{V}}
     \eqno(10)
$$
We have thus arrived at an ansatz that, while it might not be unique,
succeeds in producing the kind of fluctuations we were seeking. 
Moreover, it lends itself nicely to simulation by computer.

\subsection{Numerical simulation}
Mathematically, our model is defined in the first place by eq. (9a),
and secondly by the ansatz for $\Lambda$ described above, according to
which 
$S=V \Lambda$ is the sum of $N=V/\ell^4$ independent random
contributions, where $V$ is the volume of the spacetime region within
the past light cone of any point in the hypersurface on which $\Lambda$
is being evaluated.
Strictly speaking, this scheme is not consistently defined since it
mixes discrete variables with derivatives; however with $N\sim 10^{240}$
elements currently to our past, we are so close to the
continuum that we can safely treat our model as defined by a pair
of stochastic differential equations,
$$
    {da \over a} = \sqrt{\rho+\Lambda\over 3} d\tau  \eqno(11)
$$
$$
   V d \Lambda = V d(S/V) = dS - \Lambda \dot{V} d\tau \eqno(12)
$$
where (11) is just a rewriting of (9a).
(Perhaps, though, we should call these integro-differential
equations, inasmuch as $V$, and therefore also the ``stochastic driving
term'' $dS$, depends on the whole past-history of $a(\tau)$.)

At this point
we could try to give our equations a more precise mathematical meaning, 
but it is just as easy to pass directly to a finite-difference
form of them suitable for a computer. 
I suspect that the scheme described below 
corresponds to
the so-called It{\^o} form of the stochastic system
(11)-(12).  If so, one might consider also the Stratonovich
alternative, but we have not done so.

In references [2] and [3] our model was
simulated as follows.  Let $a_i$ be the cosmic scale-factor at the $i$th
step and similarly for $N_i$, $V_i$, $S_i$ and $\Lambda_i$.
Let $\rho=\rho_{matter}+\rho_{radition}$, with $\rho_{matter}$ taken to
be ``dust'' scaling like $1/a^3$.  
Begin at the ``Planck time'' with the appropriate ratio of
$\rho_{matter}/\rho_{radition}$ to end up matching our current universe.
Then evolve by iterating the following steps.

$\bullet$ \qquad  $a_{i+1} = a_i + a_i \sqrt{\rho_i+ \Lambda_i\over 3} (\tau_{i+1}-\tau_i)$

$\bullet$ \qquad Given $a_{i+1}$, compute $V_{i+1}$ using\footnote{$^\flat$}
{By rearranging the formulas, one can avoid recomputing this whole
 integral at each iteration. }
$$
   V(\tau) = {4\pi\over 3}\int_0^\tau d\tau' a(\tau')^3 
   \left(\int_{\tau'}^{\tau} {d\tau''\over a(\tau'')}\right)^3
$$

$\bullet$ \qquad $N_{i+1} = V_{i+1} / \ell^4$

$\bullet$ \qquad  $S_{i+1} = S_i + \xi \sqrt{N_{i+1}-N_i}$  
   \quad ($\xi$ gaussian with unit variance)

$\bullet$ \qquad  $\Lambda_{i+1} = S_{i+1}/V_{i+1}$

\noindent
(The random variable $\xi$ is gaussian thanks to the central limit
theorem. It has unit variance because our ansatz arbitrarily took each
causet element to contribute $\pm\hbar=\pm1$ to $S$.  More justified
would be $\pm\sigma$ with $\sigma$ of order unity, but we may omit this
new parameter since it coalesces with $\ell$ in its effect on the model.)

An extensive discussion of the simulations can be found
in [3] and [2].  The most important finding
was that ``tracking behavior'' was indeed observed: the
absolute value of $\Lambda$ follows $\rho_{radiation}$ very closely
during the era of radiation dominance, and then follows $\rho_{matter}$
when the latter dominates.
Secondly, the simulations confirmed that
$\Lambda$ fluctuates with a ``coherence time'' which is $O(1)$
relative to the Hubble scale.
Thirdly, a range of present-day values of $\Omega_\Lambda$ is produced,
and these are $O(1)$ when $\ell^2=O(\kappa)$.  
(Notice in this connection that the variable $\Lambda$ of our model
 cannot simply be equated to the observational parameter $\Lambda^{obs}$
 that gets reported on the basis of supernova observations, for example,
 because $\Lambda^{obs}$ results from a fit to the data that presupposes
 a constant $\Lambda$, or if not constant then a deterministically
 evolving $\Lambda$ with a simple ``equation of state''.  It turns out
 that correcting for this tends to make large values of $\Omega_\Lambda$
 more likely [3].)
Fourthly, the $\Lambda$-fluctuations affect the age of the cosmos (and
the horizon size), but not too dramatically.  
In fact they tend to increase it more often than not.  
Finally, the choice of (9a) for our specific model 
seems to be ``structurally stable`` in the sense that the results remain
qualitatively unchanged if one replaces (9a) by some linear
combination thereof with (9b), as discussed above.

I should emphasize though, that all these results come from simulations with
the free parameter $\ell$ closer to  3 than to 1 (ie to $\sqrt\kappa$).
If one lowers $\ell$ much beyond this, the negative fluctuations in
$\Lambda$ typically grow so large that the simulation cannot continue
long enough 
for the cosmos to reach its present size.
(The square root in (11) becomes imaginary.
In the ``linear combination'' version of the model, the evolution can
continue, but the cosmos recollapses to a singularity.)
This creates a tension in the model between the need for the cosmos to
reach its present size and the need for the present value of 
$\Omega_\Lambda=\Lambda/3H^2$ to be as big as it seems to be.

\subsection{Summary and Outlook}
Heuristic reasoning rooted in the basic hypotheses of causal set theory
predicted $\Lambda\sim\pm1/\sqrt{V}$, in agreement with current data.
But a fuller understanding of this prediction awaits the ``new QCD''
(``quantum causet dynamics'').  Meanwhile, a reasonably coherent
phenomenological model exists, based on simple general arguments.  It is
broadly consistent with observations but a fuller comparison is needed.
It solves the ``why now'' problem: $\Lambda$ is ``ever-present''.  
It predicts further that $p_\Lambda\not=-\rho_\Lambda$ ($w\not=-1$) and
that $\Lambda$ has probably changed its sign many times in the
past.\footnote{$^\star$} 
{It also tends to favor the existence of something, say a ``sterile
 neutrino'', to supplement the energy density at
 nucleosynthesis time.  Otherwise, we might have to assume that
 $\Omega_\Lambda$ had fluctuated to an unusually small value at that
 time.
 It also carries the implication that ``large extra dimensions'' will not be
 observed at the LHC [8]. }
The model contains a single free parameter of order unity that must be
neither too big nor too small.\footnote{$^\dagger$}
{unless we want to try to make sense of imaginary time (= quantum
 tunneling?)  or to introduce new effects to keep the right hand side of
 (9) positive (production of gravitational waves?  onset of
 large-scale spatial curvature or ``buckling''?).}
In principle the value of this parameter is calculable, but for now it
can only be set by hand.

In this connection, it's intriguing that there exists an analog
condensed matter system the ``fluid membrane'', whose analogous
parameter is not only calculable in principle from known physics, but
might also be measurable in the laboratory! [9]

That our model so far presupposes spatial homogeneity and isotropy
is no doubt its weakest feature.  Indeed, the ansatz on which it is
based strongly suggests a generalization such that
$\Lambda$-fluctuations in ``causally disconnected'' regions would be
independent of each other; and in such a generalization, spatial
inhomogeneities would inevitably arise.  Such inhomogeneities were a
source of worry in [2] and their potential to disagree
badly with the isotropy of the CMB brightness has recently been
emphasized in [10] and [11].
On the other hand they could also act as a new type of source for
density fluctuations.  Without a generalized model allowing for
spatial inhomogeneities, one cannot do better than guessing.
Let me just note that such a model would evidently have to replace the
Einstein equation by some sort of ``stochastic PDE'', just as our
homogeneous model led to a stochastic form of the Friedmann
equations.

\subsection{Closing remarks}
In itself the smallness of $\Lambda$ is a riddle and not a
problem.  But in a fundamentally discrete theory, recovery of the
continuum {\it is} a problem, and I think that the solution of this
problem will also explain the smallness of $\Lambda$. 
(The reason is that if $\Lambda$ were to take its ``natural'',
Planckian value, 
the radius of curvature of spacetime would
also be Planckian, but 
in a discrete theory
such a spacetime could no more make sense
than
a sound wave with a wavelength smaller than the size of an atom.
Therefore the only kind of spacetime that can emerge from a causet or
other discrete structure is one with $\Lambda{\ll}1$.)
One can also give good reasons why the emergence of a manifold from a
causet must rely on some form of nonlocality.
The size of $\Lambda$ should also be determined nonlocally then, and
this is precisely the kind of idea realized in the above model.

One pretty consequence of this kind of nonlocality is a certain
restoration of symmetry between the very small and the very big.
Normally, we think of $G$ (gravity) as important on large scales, with
$\hbar$ (quantum) important on small ones.  But we also expect that on
still smaller scales $G$ regains its importance once again 
and shares it with $\hbar$
(quantum
gravity).  If the
concept of an
 ever-present $\Lambda$ is correct then symmetry is
restored, because $\hbar$ rejoins $G$ on the largest scales in connection with
the cosmological constant.

Finally, let me mention a ``fine tuning'' that our model has not done
away with, namely the tuning of the spacetime dimension to $d=4$.  In
any other dimension but 4, 
$\Lambda$ could not be ``ever-present'', or rather
it could not remain in balance with matter.  
Instead, the same
crude estimates that above led us to expect $\Lambda\sim H^2$, 
lead us
in other dimensions to expect either matter dominance ($d>4$) or
$\Lambda$-dominance ($d<4$).  Could this be a dynamical reason
favoring $3+1$ as the number of noncompact dimensions?
[12] [10]

\subsection{A last word}

The cosmological constant is just as constant as Hubble's constant.



\bigskip
\noindent
It's a pleasure to thank Barbara and Syd Bulman-Fleming for acquainting
me with reference [4].
Research at Perimeter Institute for Theoretical Physics is supported in
part by the Government of Canada through NSERC and by the Province of
Ontario through MRI.
This research was partly supported 
by NSF grant PHY-0404646.

\ReferencesBegin                             

\ref [1]  
 Rafael D.~Sorkin, ``A Modified Sum-Over-Histories for Gravity'',
   reported in
  {\it 
   Highlights in gravitation and cosmology: 
   Proceedings of the International Conference on Gravitation and Cosmology, 
   Goa, India, 14-19 December, 1987},
   edited by 
   B.~R. Iyer, Ajit Kembhavi, Jayant~V. Narlikar, and C.~V. Vishveshwara,
   see pages 184-186 in the article by 
   D.~Brill and L.~Smolin: 
   ``Workshop on quantum gravity and new directions'', pp 183-191 
   (Cambridge University Press, Cambridge, 1988).
\sepref
 --------- , 
``On the Role of Time in the Sum-over-histories Framework for Gravity'',
    paper presented to the conference on 
    The History of Modern Gauge Theories, 
    held Logan, Utah, July 1987, 
    published in  
    \journaldata {Int. J. Theor. Phys.}{33}{523-534}{1994}.
\sepref
 --------- , 
``First Steps with Causal Sets'', 
  in R. Cianci, R. de Ritis, M. Francaviglia, G. Marmo, C. Rubano, 
     P. Scudellaro (eds.), 
  {\it General Relativity and Gravitational Physics} 
   (Proceedings of the Ninth Italian Conference of the same name, 
     held Capri, Italy, September, 1990), pp. 68-90
  (World Scientific, Singapore, 1991).
\sepref
 --------- , 
``Forks in the Road, on the Way to Quantum Gravity'', talk 
   given at the conference entitled ``Directions in General Relativity'',
   held at College Park, Maryland, May, 1993,
   published in
   \journaldata{Int. J. Th. Phys.}{36}{2759--2781}{1997}   ,
   \eprint{gr-qc/9706002} ,
   \lbr
   \webhomePapers
   \lbr
   (Syracuse University Preprint: SU-GP-93-12-2).
\sepref
 --------- , 
  ``Discrete Gravity'',
  a series of lectures to the 
  {\it First Workshop on Mathematical Physics and Gravitation},
   held Oaxtepec, Mexico, Dec. 1995
   (unpublished).
   %
\sepref
Y.~Jack Ng and H.~van Dam, ``A small but nonzero cosmological constant'',
\journaldata{Int. J. Mod. Phys D.}{10}{49}{2001}
\eprint{hep-th/9911102}.
\sepref

\ref [2]	
Maqbool Ahmed, Scott Dodelson, Patrick Greene and Rafael D.~Sorkin,
``Everpresent $\Lambda$'',
\journaldata {Phys. Rev.~D} {69} {103523} {2004},
\eprint{astro-ph/0209274},
\lbr
\webhomePapers.

\ref [3] 
Maqbool Ahmed,
Doctoral dissertation (Syracuse University, 2006)
\sepref
Maqbool Ahmed and Rafael D.~Sorkin,
``Everpresent Lambda II: Structural Stability and Better Comparison with Observations''
(in preparation)

\ref [4]		
Andrew McKellar,		
``The problem of possible molecular identification for interstellar lines'', 
 \journaldata{Publications of the Astronomical Society of the Pacific}{53}{233-235}{1941}
  \eprint{http://adsbit.harvard.edu/cgi-bin/nph-iarticle\_query?1941PASP...53..233M}

\ref [5]		
Luca Bombelli, Joohan Lee, David Meyer and Rafael D.~Sorkin, 
``Spacetime as a Causal Set'', 
  \journaldata {Phys. Rev. Lett.}{59}{521-524}{1987}.
\sepref
Rafael D.~Sorkin,
``Causal Sets: Discrete Gravity (Notes for the Valdivia Summer School)'',
in {\it Lectures on Quantum Gravity}
(Series of the Centro De Estudios Cient{\'\i}ficos),
proceedings of the Valdivia Summer School, 
held January 2002 in Valdivia, Chile, 
edited by Andr{\'e}s Gomberoff and Don Marolf 
(Plenum, 2005)
\eprint{gr-qc/0309009}
\sepref
Fay Dowker, ``Causal sets and the deep structure of Spacetime'', 
 in
 {\it 100 Years of Relativity - Space-time Structure: Einstein and Beyond}"
 ed Abhay Ashtekar 
 (World Scientific, to appear)
 \eprint{gr-qc/0508109}.
\sepref
Joe Henson, ``The causal set approach to quantum gravity''
 \eprint{gr-qc/0601121}.

\ref [6]		
David P.~Rideout and Rafael D.~Sorkin,
``A Classical Sequential Growth Dynamics for Causal Sets'',
 \journaldata{Phys. Rev.~D}{61}{024002}{2000}
 \eprint{gr-qc/9904062}.
\sepref
Madhavan Varadarajan and David Rideout,		
``A general solution for classical sequential growth dynamics of Causal Sets''
\journaldata {Phys. Rev. D} {73} {104021} {2006}
\eprint{gr-qc/0504066}.

\ref [7]  
M.~Henneaux and C.~Teitelboim,  
``The Cosmological Constant and General Covariance'', 
  \journaldata {Phys. Lett. B} {222} {195} {1989}

\ref [8] 
Rafael D.~Sorkin,
``Big extra dimensions make $\Lambda$ too small'',
 in the proceedings of the 
 Second International Workshop ``DICE 2004'' 
 held September, 2004, Piombino, Italy,
 edited by Hans-Thomas Elze
\eprint{gr-qc/0503057}
\journaldata{Brazilian Journal of Physics}{35}{280-283}{2005}

\ref [9]		
Joseph Samuel, Supurna Sinha, 
``Surface Tension and the Cosmological Constant''
\journaldata{Phys. Rev. Lett.}{97}{161302}{2006},
\eprint{arXiv:cond-mat/0603804} 

\ref [10]   
John D. Barrow,
``A Strong Constraint on Ever-Present Lambda'',
\journaldata{Phys.Rev. D}{75}{067301}{2007},
\eprint{arXiv:gr-qc/0612128}

\ref [11] Joe Zuntz,
"The CMB in a Causal Set Universe"
(in preparation)

\ref [12]
Rafael D.~Sorkin,
``Why the Cosmological Constant can't quite vanish'',
Talk delivered at the Third Eastern Gravity Meeting, 
held {March 1999} at Cornell University, Ithaca, NY.

\end               


(prog1    'now-outlining
  (Outline 
     "\f......"
      "
      "
      "
   ;; "\\\\message"
   "\\\\Abstrac"
   "\\\\section"
   "\\\\subsectio"
   "\\\\appendi"
   "\\\\Referen"
   "\\\\ref....[^|]"
  ;"\\\\ref....."
   "\\\\end